\documentclass[12pt]{article}
\usepackage{latexsym}
\usepackage{amsmath,amsfonts}
\usepackage{times}
\allowdisplaybreaks[4]

\newcommand{\No}{\mathbb{N}}

\catcode`\@=11

\newcount\hour
\newcount\minute
\newtoks\amorpm \hour=\time\divide\hour by 60\minute
=\time{\multiply\hour by 60 \global\advance\minute by-\hour}
\edef\standardtime{{\ifnum\hour<12 \global\amorpm={am}%
        \else\global\amorpm={pm}\advance\hour by-12 \fi
        \ifnum\hour=0 \hour=12 \fi
        \number\hour:\ifnum\minute<10
        0\fi\number\minute\the\amorpm}}
\edef\militarytime{\number\hour:\ifnum\minute<10 0\fi\number\minute}

\def\draftlabel#1{{\@bsphack\if@filesw {\let\thepage\relax
   \xdef\@gtempa{\write\@auxout{\string
      \newlabel{#1}{{\@currentlabel}{\thepage}}}}}\@gtempa
   \if@nobreak \ifvmode\nobreak\fi\fi\fi\@esphack}
        \gdef\@eqnlabel{#1}}
\def\@eqnlabel{}
\def\@vacuum{}
\def\marginnote#1{}
\def\draftmarginnote#1{\marginpar{\raggedright\scriptsize\tt#1}}
\def\draft{
        \pagestyle{plain}
        \overfullrule=2pt
        \oddsidemargin -.5truein
        \def\@oddhead{\sl \phantom{\today\quad\militarytime} \hfil
        \smash{\Large\sl DRAFT} \hfil \today\quad\militarytime}
        \let\@evenhead\@oddhead
        \let\label=\draftlabel
        \let\marginnote=\draftmarginnote
        \def\ps@empty{\let\@mkboth\@gobbletwo
        \def\@oddfoot{\hfil \smash{\Large\sl DRAFT} \hfil}
        \let\@evenfoot\@oddhead}
        \def\@eqnnum{(\theequation)\rlap{\kern\marginparsep\tt\@eqnlabel}%
        \global\let\@eqnlabel\@vacuum}  }

\newcommand{\rf}[1]{(\ref{#1})}
\renewcommand{\theequation}{\thesection.\arabic{equation}}
\renewcommand{\thefootnote}{\fnsymbol{footnote}}
\newcommand{\newsection}{   
\setcounter{equation}{0}\section}

\def\appendix#1{\addtocounter{section}{1}\setcounter{equation}{0}
\renewcommand{\thesection}{\Alph{section}}
\section*{Appendix \thesection\protect\indent \parbox[t]{11.15cm}{#1}}
\addcontentsline{toc}{section}{Appendix \thesection\ \ \ #1}}

\def\be{\begin{equation}}
\def\ee{\end{equation}}
\def\beq{\begin{eqnarray}}
\def\eeq{\end{eqnarray}}

\def\parline{\,\partial\kern -0.55em /\,\,}

\def\half{{\frac{1}{2}}}

\def\AA{{\cal A}}

\def\GG{{\cal G}}

\def\LL{{\cal L}}

\def\gb{\bar{g}}

\def\rb{\bar{r}}

\def\psik{|\psi\rangle}
\def\psibr{\langle\psi|}

\def\xik{|\xi\rangle}

\def\smG{{\scriptscriptstyle G}}

\def\smzero{{\scriptscriptstyle (0)}}
\def\smone{{\scriptscriptstyle (1)}}

\def\oplussm{{\scriptscriptstyle \oplus}}
\def\ominussm{{\scriptscriptstyle \ominus}}

\def\Gammasm{{\scriptscriptstyle \Gamma}}
\def\Esm{{\scriptscriptstyle E}}
\def\Gsm{{\scriptscriptstyle G}}
\def\smE{{\scriptscriptstyle E}}
\def\smGE{{\scriptscriptstyle GE}}

\def\smG{{\scriptscriptstyle G}}

\def\smzero{{\scriptscriptstyle (0)}}
\def\smone{{\scriptscriptstyle (1)}}

\def\oplussm{{\scriptscriptstyle \oplus}}
\def\ominussm{{\scriptscriptstyle \ominus}}

\def\smponetwo{{\scriptscriptstyle [1,2]}}
\def\smponethree{{\scriptscriptstyle [1,3]}}

\def\rwt{\widetilde{r}}

\def\gaal{\gamma\alpha}
\def\gaalb{\gamma\bar\alpha}

\def\diff{{\rm diff}}

\def\u{{\rm u}}
\def\d{{\rm d}}

\def\ibf{{\bf i}}
\def\iibf{{\bf ii}}
\def\iiibf{{\bf iii}}
\def\ivbf{{\bf iv}}

\def\nbf{{\bf n}}

\begin{document}


\begin{flushright}
FIAN-TD-2012-35 \hspace{1.7cm}{}~\\
arXiv: 1211.4498 [hep-th] \hspace{0.5cm}{}~\\
Updated September 2019 \hspace{0.6cm}{}~
\end{flushright}

\vspace{1cm}

\begin{center}

{\Large \bf Conformal totally symmetric arbitrary spin

\bigskip  fermionic fields}

\vspace{2.5cm}

R.R. Metsaev\footnote{ E-mail: metsaev@lpi.ru }

\vspace{1cm}

{\it Department of Theoretical Physics, P.N. Lebedev Physical
Institute,
\\ Leninsky prospect 53,  Moscow 119991, Russia }

\vspace{3.5cm}

{\bf Abstract}

\end{center}

Conformal totally symmetric arbitrary spin fermionic fields propagating in the flat
space-time of even dimension greater than or equal to four are
investigated. First-derivative metric-like formulation involving Fang-Fronsdal
kinetic operator for such fields is developed. Gauge invariant
Lagrangian and the corresponding gauge transformations are obtained.
Gauge symmetries of the Lagrangian are realized by using auxiliary fields and the Stueckelberg fields. Realization of conformal algebra symmetries on the
space of conformal gauge fermionic fields is obtained. The on-shell
degrees of freedom of the fermionic arbitrary spin conformal fields
are also studied.

\newpage
\renewcommand{\thefootnote}{\arabic{footnote}}
\setcounter{footnote}{0}

\newsection{\large Introduction}

Although, at present time, up-to date methods of quantization of gauge theories allow to treat  higher-derivative theories and theories that do not consist higher derivatives on an equal footing, we note that the use of the famous Slavnov-Taylor identities \cite{Slavnov:1972fg} and BRST approach \cite{Becchi:1975nq} is streamlined when the gauge fields theories do not consist higher derivatives. Note however that commonly used Lagrangian formulations of most conformal fields
consist higher derivatives \cite{Fradkin:1985am}. We recall that Lagrangian higher-derivative formulation
of {\it bosonic} totally symmetric arbitrary spin conformal
fields in $R^{d,1}$ space for $d=4$ and $d\geq 4$ was obtained in the respective Ref.\cite{Fradkin:1985am} and
Ref.\cite{Segal:2002gd}.%
\footnote{ Higher-derivative Lagrangian formulation of mixed-symmetry conformal bosonic
fields was obtained in Ref.\cite{Vasiliev:2009ck}. Discussion of
equations of motions and conservation laws for mixed-symmetry conformal fields may be
found in Refs.\cite{Shaynkman:2004vu,Dobrev:2007vn,Alkalaev:2012rg}).}
Alternative higher-derivative formulation of  {\it
bosonic} conformal fields obtained by using AdS/CFT correspondence was developed in
Ref.\cite{Metsaev:2009ym}. At present time, higher-derivative Lagrangian formulation
of the totally symmetric arbitrary spin conformal {\it fermionic}
is known only for the case of $R^{3,1}$ space (see Ref.\cite{Fradkin:1985am}).

The present paper is a continuation of our investigations in Refs.\cite{Metsaev:2007fq,Metsaev:2007rw}. In Ref.\cite{Metsaev:2007fq}, for free {\it bosonic and fermionic low-spin} conformal
fields, we developed the respective second-derivative and first-derivative Lagrangian gauge invariant  metric-like formulations. The second-derivative metric-like formulation of {\it bosonic arbitrary
spin} conformal fields was developed in Ref.\cite{Metsaev:2007rw}.
In this paper, we develop Lagrangian gauge invariant metric-like formulation for totally symmetric {\it fermionic arbitrary spin} fields in $R^{d-1,1}$ space, $d \geq
4$.

Our approach to conformal fermionic fields is summarized as follows.

\noindent \ibf) In addition to fields entering the higher-derivative formulation of conformal fields we introduce Stueckelberg fields and auxiliary fields.

\noindent \iibf) Kinetic term of our Lagrangian of conformal fermionic field does not involve higher than first order terms in the derivatives. The one-derivative contributions to the
kinetic terms of Lagrangian of conformal fermionic fields are realized as the well-known Dirac, Rarita-Schwinger, and Fang-Fronsdal kinetic terms of the respective spin-$\half$,
spin-$\frac{3}{2}$, and spin-$(s+\half)$, $s>1$, $s\in \No$,
fermionic fields.

\noindent \iiibf) All vector-spinor and tensor-spinor fermionic
fields entering our Lagrangian are supplemented
by the respective gauge symmetries.%
\footnote{ In our approach, we realize gauge symmetries by adapting the
approach in Refs.\cite{Zinoviev:2001dt,Metsaev:2006zy}.}
Gauge transformations of the fermionic fields do not contain
higher than first order terms in derivatives. One-derivative
contributions to gauge transformations of fermionic conformal fields are realized as the
well-known gradient gauge transformations of the vector-spinor and
tensor-spinor fermionic fields.

\noindent \ivbf) Our first-derivative formulation is equivalent to the higher-derivative formulation. Namely, by eliminating the auxiliary fields via equations of motion and gauging away the Stueckelberg fields, we can verify that our first-derivative formulation of conformal fermionic fields amounts to the higher-derivative formulation of conformal fermionic fields.

The rest of the paper  is organized as follows.

In Sec. \ref{prelim},  we summarize our notation.
In Sec.\ref{lagran}, we develop the first-derivative metric-like formulation
for arbitrary spin conformal fermionic field. In
Sec.\ref{fieldcont}, we start with the discussion of field content
entering our approach. After this, in Sec.\ref{gaugsym}, we present
our result for our first-derivative gauge invariant Lagrangian
and realization of gauge symmetries in our approach. In
Sec.\ref{realcondsymmetr}, we discuss realization of conformal
algebra symmetries on the space of gauge fields entering our
approach. In Sec.\ref{dof}, we describe our results for number
of on-shell degrees of freedom (D.o.F) for the arbitrary spin conformal fermionic field
and decomposition of those on-shell D.o.F into irreps of the
$so(d-2)$ algebra. Section \ref{conlcus} is devoted to the
discussion of directions for future research.

\newsection{\large Notation and conventions} \label{prelim}

Our conventions are as follows. $x^a$ denotes coordinates in the
$R^{d-1,1}$ space, while $\partial_a$ denotes
derivatives with respect to $x^a$, $\partial_a \equiv \partial /
\partial x^a$. Vector indices of the Lorentz algebra $so(d-1,1)$ take
the values $a,b,c,e=0,1,\ldots ,d-1$. To simplify our expressions we drop the flat metric
$\eta_{ab}=(-,+,\ldots, +)$ in scalar products, i.e., we use $X^aY^a \equiv
\eta_{ab}X^a Y^b$.

A set of the creation operators $\alpha^a$, $\zeta$,
$\upsilon^\oplussm$, $\upsilon^\ominussm$ and the respective set of
annihilation operators $\bar{\alpha}^a$, $\bar\zeta$, $\bar\upsilon^\oplussm$, $\bar\upsilon^\ominussm$ will be referred
to as oscillators in what follows.%
\footnote{ We use the oscillator to introduce generating ket-vectors and simplify our expressions (see also Refs.\cite{Vasiliev:1987tk,Hallowell:2005np,Bekaert:2006ix}).}
Commutation relations and the vacuum are defined as
\beq
&&{} [\bar{\alpha}^a,\alpha^b]=\eta^{ab}\,, \qquad
[\bar\zeta,\zeta]=1\,, \qquad [\bar{\upsilon}^\oplussm,\,
\upsilon^\ominussm ]=1\,, \qquad\quad [\bar{\upsilon}^\ominussm,\,
\upsilon^\oplussm]=1\,,
\\
&& \bar\alpha^a |0\rangle = 0\,,\qquad\quad  \bar\zeta|0\rangle =
0\,,\qquad\quad \bar\upsilon^\oplussm |0\rangle = 0\,,\qquad\quad
\quad \bar\upsilon^\ominussm |0\rangle = 0\,.
\eeq
The oscillators $\alpha^a$, $\bar\alpha^a$ and $\zeta$, $\bar\zeta$,
$\upsilon^\oplussm$, $\upsilon^\ominussm$, $\bar\upsilon^\oplussm$,
$\bar\upsilon^\ominussm$ transform in the respective vector and
scalar representations of the Lorentz algebra $so(d-1,1)$. We use
$2^{[d/2]}\times 2^{[d/2]}$ Dirac gamma matrices $\gamma^a$, $ \{ \gamma^a,\gamma^b\} = 2\eta^{ab}$, and adapt
the following hermitian conjugation rules for the derivatives,
oscillators, and $\gamma$-matrices:
\be \label{03082011-01} \partial^{a\dagger} = - \partial^a, \quad
\gamma^{a\dagger} = \gamma^0 \gamma^a\gamma^0\,,\quad
\alpha^{a\dagger} = \bar\alpha^a\,, \quad \zeta^\dagger = \bar\zeta
\,,
\quad
\upsilon^{\oplussm\dagger} = \bar\upsilon^\oplussm\,,\quad
\upsilon^{\ominussm \dagger} = \bar\upsilon^\ominus \,.
\ee
We use operators constructed out of the derivatives, oscillators,
and Dirac $\gamma$-matrices,
\beq \label{manold-31102011-02}
&& \Box \equiv \partial^a\partial^a\,, \qquad\quad
\parline\equiv \gamma^a\partial^a\,,\qquad\quad
\alpha\partial \equiv \alpha^a\partial^a\,,\qquad\quad
\bar\alpha\partial \equiv \bar\alpha^a\partial^a\,,\qquad
\\
\label{manold-31102011-03} && \gamma\alpha \equiv
\gamma^a\alpha^a\,,\qquad\ \ \gamma\bar\alpha \equiv
\gamma^a\bar\alpha^a\,,\qquad \ \ \
\alpha^2 \equiv \alpha^a\alpha^a\,, \qquad\quad   \bar\alpha^2
\equiv \bar\alpha^a\bar\alpha^a\,,\qquad
\\
\label{manold-31102011-05} && N_\alpha \equiv \alpha^a \bar\alpha^a
\,,
\qquad \
N_\zeta \equiv \zeta \bar\zeta \,, \qquad\quad \ \ \
N_{\upsilon^\oplussm} \equiv \upsilon^\oplussm
\bar\upsilon^\ominussm\,, \qquad \
N_{\upsilon^\ominussm} \equiv \upsilon^\ominussm
\bar\upsilon^\oplussm\,,
\\
\label{manold-31102011-06} && N_\upsilon \equiv
N_{\upsilon^\oplussm} + N_{\upsilon^\ominussm}\,,
\qquad\qquad\qquad \quad \ \
\Delta' \equiv N_{\upsilon^\ominussm} - N_{\upsilon^\oplussm} \,.
\\
&& \AA^a \equiv \alpha^a - \gaal \gamma^a \frac{1}{2N_\alpha+d-2} -
\alpha^2 \frac{1}{2N_\alpha+d} \bar\alpha^a\,,
\\
&& \GG^a \equiv \gamma^a - \gaal \frac{2}{2N_\alpha+d-2}
\bar\alpha^a\,,
\\
&& \AA_1^a \equiv \alpha^a - \gaal \gamma^a \frac{1}{2N_\alpha+d} -
\alpha^2 \frac{1}{2N_\alpha+d+2} \bar\alpha^a\,,
\\
&& \bar\AA_1^a \equiv \alpha^a - \frac{1}{2N_\alpha+d}
\gamma^a\gaalb - \alpha^a \frac{1}{2N_\alpha+d+2} \bar\alpha^2\,,
\\
&& \GG_1^a \equiv \gamma^a - \gaal \frac{2}{2N_\alpha+d}
\bar\alpha^a\,,
\\
&& \bar\GG_1^a \equiv \gamma^a - \alpha^a \frac{2}{2N_\alpha+d}
\gaalb\,,
\\
&& \Pi^\smponethree = 1 - \gaal \frac{1}{2N_\alpha+d}\gaalb -
\alpha^2 \frac{1}{2(2N_\alpha +d+2)}\bar\alpha^2\,,
\\
&& \Pi^\smponetwo = 1 - \alpha^2 \frac{1}{2(2N_\alpha
+d)}\bar\alpha^2\,.
\eeq
The $2\times 2$ matrices $\sigma_\pm$, $\sigma_3$, $\pi_\pm$, and
antisymmetric products of $\gamma$-matrices are defined as
\beq \label{manold-03112011-01}
&& \sigma_+  = \left(
\begin{array}{ll}
0 & 1
\\
0 & 0
\end{array}\right),
\quad
\sigma_-  = \left(
\begin{array}{ll}
0 & 0
\\
1 & 0
\end{array}\right),
\quad
\sigma_3  = \left(
\begin{array}{lc}
1 & 0
\\
0 & -1
\end{array}\right),
\quad
\pi_\pm   = \half (1 \pm \sigma_3)\,,\qquad
\\
\label{manold-31102011-04} && \hspace{1cm} \gamma^{ab} = \half
(\gamma^a \gamma^b - \gamma^b \gamma^a)\,,\qquad \gamma^{abc} =
\frac{1}{3!}(\gamma^a\gamma^b\gamma^c \pm 5 \hbox{ terms})\,.
\eeq
The notation $k' \in [n]_2$ implies that $k' =-n,-n+2,-n+4,\ldots,n-4, n-2,n$:
\be
\label{sumnot02} k' \in [n]_2 \quad \Longleftrightarrow \quad k'
=-n,-n+2,-n+4,\ldots,n-4, n-2,n\,.
\ee
Using the notation  $A^\dagger$ for the standard hermitian conjugated of
the operator $A$ we define operator $A^{\widehat{\dagger}}$ as follows
\be
A^{\widehat{\dagger}} \equiv  - \gamma^0 A^\dagger \gamma^0\,.
\ee

\newsection{ \large First-derivative gauge
invariant Lagrangian}\label{lagran}

\subsection{Field content}\label{fieldcont}

In order to develop the first-derivative gauge invariant metric-like formulation of
spin-$(s+\half)$ conformal non-chiral Dirac fermionic field in $R^{d-1,1}$ space, $d\geq 4$, we use the set of non-chiral spinor,
vector-spinor, and tensor-spinor Dirac fields of the Lorentz algebra
$so(d-1,1)$ given by:
\beq
&& \hspace{-0.5cm} \psi_{k'}^{a_1\ldots a_{s'}} \,,\hspace{1.7cm}
s'=0,1,\ldots,s\,, \hspace{5.3cm} k' \in [k_{s'}]_2;
\nonumber\\[-10pt]
\label{man02-12112012-01} &&
\\[-10pt]
&& \hspace{-0.5cm} \psi_{k'}^{a_1\ldots a_{s'}}\,, \hspace{1.7cm}
s'= \left\{\begin{array}{l}
0,1,\ldots,s;\qquad  \hbox{for }\  d \geq 6;
\\[3pt]
1,2,\ldots,s;\qquad  \hbox{for }d = 4;
\end{array}\right.
\hspace{2cm} k' \in  [k_{s'}-1]_2\,;\qquad
\nonumber\\
\label{man02-12112012-03} && \hspace{3.6cm} k_{s'} \equiv s'+
\frac{d-4}{2} \,,
\eeq
where the spinor indices of the fermionic fields
$\psi_{k'}^{a_1\ldots a_{s'}}$ are implicit and for some notation
see \rf{sumnot02}. The fields $\psi_{k'}^{a_1\ldots a_{s'}}$ with $s'=0$, $s'=1$, and $s'\geq 2$, are the the respective
non-chiral spinor, vector-spinor, and tensor-spinor fermionic fields
of the Lorentz algebra $so(d-1,1)$. Chiral fermionic fields
are discussed below.

For $d\geq 6$, field content given in \rf{man02-12112012-01}
can alternatively be represented as
\begin{eqnarray}
\label{man02-12112012-04} &&\psi_{k'}^{a_1\ldots a_s}\,,
\hspace{3cm} k'\in [k_s]_2\,;
\\
&&\psi_{k'}^{a_1\ldots a_s}\,, \hspace{3cm} k'\in [k_s-1]_2\,;
\\
&&\psi_{k'}^{a_1\ldots a_{s-1}}\,, \hspace{2.7cm} k' \in
[k_{s-1}]_2\,;
\\
&&\psi_{k'}^{a_1\ldots a_{s-1}}\,, \hspace{2.7cm} k' \in
[k_{s-1}-1]_2\,;
\\
&&\ldots\ldots \hspace{3.4cm} \ldots \ldots\ldots
\nonumber\\[-10pt]
&&\ldots\ldots \hspace{3.4cm} \ldots \ldots\ldots
\nonumber\\
&& \psi_{k'}^a\,, \hspace{3.8cm} k' \in [k_1]_2\,;
\\
&& \psi_{k'}^a\,, \hspace{3.8cm} k' \in [k_1-1]_2\,;
\\
\label{man02-12112012-10} && \psi_{k'}\,, \hspace{3.8cm} k' \in
[k_0]_2\,;
\\
\label{man02-12112012-11} && \psi_{k'}\,, \hspace{3.8cm} k' \in
[k_0-1]_2\,;
\end{eqnarray}
while, for the case of $d=4$, field content \rf{man02-12112012-01} is given in
\rf{man02-12112012-04}-\rf{man02-12112012-10}. This is to say that
fields in \rf{man02-12112012-11} enter field content only for $d\geq
6$.

We make comments on the field content.\\
\ibf) In \rf{man02-12112012-01}, the fields $\psi_{k'}$ and
$\psi_{k'}^a$ are the respective non-chiral spinor and vector-spinor
fields of the Lorentz algebra, while the fields $\psi_{k'}^{a_1\ldots
a_{s'}}$, $s'>1$, are rank-$s'$ totally symmetric non-chiral
tensor-spinor fields of the Lorentz algebra $so(d-1,1)$.
\\
\iibf) The tensor-spinor fields $\psi_{k'}^{a_1\ldots a_{s'}}$ with
$s'\geq 3 $ satisfy the $\gamma$ triple-tracelessness constraint,
\be \label{gammatriple}
\gamma^a\psi_{k'}^{abba_4\ldots a_{s'}}=0\,, \qquad s'\geq 3\,.
\ee
\iiibf) The conformal dimension of the fermionic field $\psi_{k'}^{a_1\ldots
a_{s'}}$ is given by
\be
\label{condimarbspi01} \Delta(\psi_{k'}^{a_1\ldots a_{s'}}) =
\frac{d-1}{2} + k'\,.
\ee

In order to illustrate our field content presented in \rf{man02-12112012-01} let us
use the shortcut $\psi_{k'}^{s'}$ for the field
$\psi_{k'}^{a_1\ldots a_{s'}}$. We note then that, for $d\geq 6$ and
arbitrary $s$, fields in \rf{man02-12112012-01} can be presented as
in Table I.
\newpage
{\tiny
\begin{eqnarray*}
&& \hspace{-0.1cm} \hbox{\small {\sf TABLE I}. \  Field content for
$d \geq 6\,, s -$ arbitrary\,. The notation $\psi_{k'}^{s'}$ stands
for $\psi_{k'}^{a_1\ldots a_{s'}}.$}
\nonumber\\
&&\hspace{-1cm} \psi_{-k_s}^s \hspace{1cm} \psi_{2-k_s}^s
\hspace{1cm} \ldots \hspace{1cm}\ldots \hspace{2cm} \ldots
\hspace{2cm} \ldots \hspace{1cm}\ldots \hspace{0.9cm} \psi_{k_s-2}^s
\hspace{1cm}  \psi_{k_s}^s \qquad
\nonumber\\[15pt]
&& \hspace{-0.3cm} \psi_{1-k_s}^s \hspace{1cm} \psi_{3-k_s}^s
\hspace{1cm} \ldots \hspace{1cm} \ldots \hspace{1cm}\ldots
\hspace{1cm}\ldots \hspace{1cm}\ldots \hspace{1cm} \psi_{k_s-3}^s
\hspace{1cm} \psi_{k_s-1}^s
\nonumber\\[15pt]
&& \hspace{-0.3cm} \psi_{1-k_s}^{s-1} \hspace{1cm}
\psi_{3-k_s}^{s-1} \hspace{1cm} \ldots \hspace{1cm}\ldots
\hspace{1cm}\ldots \hspace{1cm}\hspace{1cm}\ldots \hspace{1.5cm}
\psi_{k_s-3}^{s-1} \hspace{1cm} \psi_{k_s-1}^{s-1}
\nonumber\\[15pt]
&&\hspace{-1cm} \phantom{\psi_{-k_s}^s} \hspace{1cm}
\psi_{2-k_s}^{s-1} \hspace{1cm} \ldots \hspace{1cm}\ldots
\hspace{1cm} \ldots \hspace{1cm} \ldots
\hspace{1cm}\ldots\hspace{1cm}\ldots \hspace{1.5cm}
\psi_{k_s-2}^{s-1} \hspace{1cm} \phantom{\psi_{k_s}^s} \qquad
\nonumber\\[14pt]
&& \hspace{0.8cm} \hspace{1cm}  \ldots \hspace{1cm}  \ldots
\hspace{1cm} \ldots \hspace{1cm} \ldots   \hspace{1cm} \ldots
\hspace{1cm} \ldots \hspace{1cm} \ldots
\\[14pt]
&& \hspace{1.8cm}  \psi_{-k_1+1}^1 \hspace{1cm} \psi_{-k_1+3}^1
\hspace{0.6cm}  \ldots  \hspace{0.6cm} \ldots \hspace{0.6cm}
\psi_{k_1-3}^1 \hspace{0.8cm} \psi_{k_1-1}^1
\nonumber\\[15pt]
&& \hspace{1.8cm} \psi_{-k_0}^0 \hspace{1.2cm} \psi_{-k_0+2}^0
\hspace{1.2cm} \ldots \hspace{1cm} \psi_{k_0-2}^0 \hspace{1cm}
\psi_{k_0}^0
\nonumber\\[15pt]
&& \hspace{3cm} \psi_{-k_0+1}^0 \hspace{1.2cm}  \ldots
\hspace{1.2cm} \ldots \hspace{1cm} \psi_{k_0-1}^0 \hspace{1cm}
\nonumber
\end{eqnarray*} }

The spinor fields $\psi_{k'}^0$ having $k'\in
[k_0-1]_2$ do not enter the field content when $d=4$. Namely, for $d=4$ and arbitrary $s$, the field content in
\rf{man02-12112012-01} can be represented as in Table II.
{\small
$$ \hbox{{\sf TABLE II}. \ \ Field content for} \ \ d = 4\,, \ \ s - \hbox{
arbitrary}. \hbox{ The notation $\psi_{k'}^{s'}$ stands for
$\psi_{k'}^{a_1\ldots a_{s'}}$}. $$
$$ \label{oldman-27022012-01}
\begin{array}{ccccccccccccccc}
\psi_{-s}^s & & \psi_{-s+2}^s & & &&&\ldots&&& & & \psi_{s-2}^s & &
\psi_s^s
\\[12pt]
& \psi_{-s+1}^s & & \psi_{-s+3}^s & &&&\ldots& &&& \psi_{s-3}^s & &
\psi_{s-1}^s &
\\[12pt]
& \psi_{-s+1}^{s-1} & & \psi_{-s+3}^{s-1} & &&&\ldots &&&&
\psi_{s-3}^{s-1} & & \psi_{s-1}^{s-1} &
\\[12pt]
&& \psi_{-s+2}^{s-1} & & \ldots  &   &  & \ldots && & \ldots  &  &
\psi_{s-2}^{s-1} &&
\\[12pt]
&&  &\ldots & \ldots  &   &  & \ldots && & \ldots  & \ldots &&&
\\[12pt]
&&& & \ldots  &   &  & \ldots && & \ldots  &  &&&
\\[12pt]
&&&  &   \psi_{-2}^2 &   && \psi_0^2 & &   & \psi_2^2 & &&&
\\[12pt]
&&&  &  & \psi_{-1}^2  &&& &  \psi_1^2 &  & &&&
\\[12pt]
&&&  & & \psi_{-1}^1  &&& & \psi_1^1 & & &&&
\\[12pt]
& && & &  & & \psi_0^1 & &&&  &&&
\\[12pt]
& && & &  & & \psi_0^0 & &&&  &&&
\end{array}
$$ }

We note that $d=6$ is the lowest space-time dimension when the
spinor fields $\psi_{k'}^0$ having $k'\in [k_0-1]_2$ appear in the
field content. This is to say that for the case of $d=6$ and arbitrary $s$, the field
content given in \rf{man02-12112012-01} can be represented as in Table
III.

\vspace{-0.6cm}
{\small
$$ \hbox{{\sf TABLE III}. \ \ Field content for} \ \ d = 6\,,
\ \ s - \hbox{ arbitrary}. \hbox{ The notation $\psi_{k'}^{s'}$
stands for $\psi_{k'}^{a_1\ldots a_{s'}}$}.
$$
$$
\begin{array}{ccccccccccccc}
\psi_{-s-1}^s & & \psi_{-s+1}^s & & & &\ldots && & & \psi_{s-1}^s &
& \psi_{s+1}^s
\\[12pt]
& \psi_{-s}^s & & \psi_{-s+2}^s & & &\ldots && & \psi_{s-2}^s & &
\psi_{s}^s &
\\[12pt]
& \psi_{-s}^{s-1} & & \psi_{-s+2}^{s-1} & & &\ldots & &&
\psi_{s-2}^{s-1} & & \ \ \ \ \ \psi_{s}^{s-1} &
\\[12pt]
&  & \psi_{-s+1}^{s-1} & & & &\ldots & & &  & \psi_{s-1}^{s-1} & &
\\[12pt]
& & \ldots  &  && & \ldots &&& & \ldots  &  &
\\[12pt]
& &  & \ldots && & \ldots &&& \ldots &   &  &
\\[12pt]
& &  &  & \ldots & & \ldots && \ldots &  &   &  &

\\[12pt]
&  &  &  &\psi_{-2}^1 & & \psi_0^1 & & \psi_2^1 & &  & &
\\[12pt]
& & &  & & \psi_{-1}^1  & & \psi_1^1 & & & & &
\\[12pt]
&&&  & & \psi_{-1}^0  & & \psi_1^0 & & & & &
\\[12pt]
&&&  & &  & \psi_0^0& & & &&&
\end{array}
$$}

To illustrate further the field content entering our approach we
note that, for the case of $d=4$ and spin-$\frac{3}{2}$ field, the field
content shown in \rf{man02-12112012-01} is given by

$$ \hbox{\small {} \qquad \ Field content for $d=4$ and spin-$\frac{3}{2}$ field} $$

\vspace{-0.7cm}
\beq
&& \psi_{_{-1}}^a \qquad  \psi_{_1}^a
\nonumber\\
&& \quad \ \ \ \psi_{_0}^a
\nonumber\\
&& \quad \ \ \ \psi_{_0}
\eeq
while, for the case of $d=6$ and spin-$\frac{3}{2}$ field, the field
content in \rf{man02-12112012-01} is given by

$$ \hbox{\small {} \qquad \ Field content for $d=6$ and spin-$\frac{3}{2}$ field} $$

\vspace{-0.7cm}
\beq
&& \psi_{_{-2}}^a \qquad  \psi_{_0}^a \qquad \psi_{_{2}}^a
\nonumber\\
&& \quad \ \ \ \psi_{_{-1}}^a \qquad \psi_{_1}^a
\nonumber\\
&& \quad \ \ \ \ \psi_{_{-1}} \qquad \psi_{_1}
\nonumber\\
&& \quad \qquad \ \ \ \psi_{_0}
\eeq

For the case of $d=4$ and spin-$\frac{5}{2}$ field, the field content
in \rf{man02-12112012-01} is given by

$$ \hbox{\small {} \qquad \ Field content for $d=4$ and spin-$\frac{5}{2}$ field} $$

\vspace{-0.7cm}
\beq
&& \psi_{_{-2}}^{ab} \qquad  \psi_{_0}^{ab} \qquad \psi_{_{2}}^{ab}
\nonumber\\
&& \quad \ \ \ \psi_{_{-1}}^{ab} \qquad \psi_{_1}^{ab}
\nonumber\\
&& \quad \ \ \ \ \psi_{_{-1}}^a \qquad \psi_{_1}^a
\nonumber\\
&& \quad \qquad \ \ \ \psi_{_0}^a
\nonumber\\
&& \quad \qquad \ \ \ \psi_{_0}
\eeq
while, for the case of $d=6$ and spin-$\frac{5}{2}$ field, the field
content in \rf{man02-12112012-01} is given by
$$ \hbox{\small {} \qquad \ Field content for $d=6$ and spin-$\frac{5}{2}$ field} $$

\vspace{-0.7cm}
\beq
&&\hspace{-0.9cm} \psi_{_{-3}}^{ab} \qquad  \psi_{_{-1}}^{ab} \qquad
\psi_{_1}^{ab} \qquad \psi_{_3}^{ab}
\nonumber\\
&&{}\!\! \psi_{_{-2}}^{ab} \qquad  \psi_{_0}^{ab} \qquad
\psi_{_{2}}^{ab}
\nonumber\\
&& \psi_{_{-2}}^a \qquad  \psi_{_0}^a \qquad \ \psi_{_{2}}^a
\nonumber\\
&& \quad \ \ \ \psi_{_{-1}}^a \qquad \psi_{_1}^a
\nonumber\\
&& \quad \ \ \ \ \psi_{_{-1}} \qquad \psi_{_1}
\nonumber\\
&& \quad \qquad \ \ \ \psi_{_0}
\eeq

To simplify our presentation we use the oscillators  $\alpha^a$, $\zeta$,
$\upsilon^\oplussm$, $\upsilon^\ominussm$ to collect all fields appearing
\rf{man02-12112012-01} into the ket-vector $\psik$ given by
\beq
\label{man02-14112012-04-a} && \psik = \sum_{s'=0}^s\frac{\zeta^{s-s'}}{\sqrt{(s-s')!}}
|\psi^{s'}\rangle \,, \hspace{1cm} |\psi^{s'}\rangle  = \left(
\begin{array}{l}
|\psi_\u^{s'}\rangle
\\[7pt]
|\psi_\d^{s'}\rangle \end{array} \right),
\\
\label{man02-14112012-01} && |\psi_\u^{s'}\rangle  \equiv \sum_{k'
\in [k_{s'}]_2} \frac{ (\upsilon^\ominussm)^{^{\frac{k_{s'}+k'}{2}}}
(\upsilon^\oplussm)^{^{\frac{k_{s'}-k'}{2}}} }{
s'!(\frac{k_{s'}+k'}{2})!}\, \alpha^{a_1} \ldots  \alpha^{a_{s'}}
\psi_{ k'}^{a_1\ldots a_{s'}}|0\rangle\,,
\\
\label{man02-14112012-02} && |\psi_\d^{s'}\rangle \equiv \sum_{k'
\in [k_{s'}-1]_2} \frac{
(\upsilon^\ominussm)^{^{\frac{k_{s'}-1+k'}{2}}}
(\upsilon^\oplussm)^{^{\frac{k_{s'}-1-k'}{2}}} }{
s'!(\frac{k_{s'}-1+k'}{2})!}\, \alpha^{a_1} \ldots  \alpha^{a_{s'}}
\psi_{k'}^{a_1\ldots a_{s'}}|0\rangle\,,
\\
\label{man02-14112012-03} && \hspace{2cm} \psi_{k'} \equiv 0\,,
\qquad k'\in [k_0-1]_2\,, \qquad \hbox{ for } \ d=4\,,
\eeq
where $k_{s'}$ is given in \rf{man02-12112012-03}. For $d=4$, we use
relation \rf{man02-14112012-03} in \rf{man02-14112012-02} to respect
the fact that fields appearing in \rf{man02-12112012-11} do not
enter the field content when $d=4$. It is easy to see that
ket-vector \rf{man02-14112012-04-a} satisfies the following
algebraic constraints:
\beq
\label{man02-14112012-05} && (N_\alpha + N_\zeta - s ) \psik = 0 \,,
\\
\label{man02-14112012-06} && (N_\zeta + N_\upsilon - k_s )\pi_+
\psik = 0 \,,
\\
\label{man02-14112012-07} && (N_\zeta + N_\upsilon  - k_s+1)\pi_-
\psik =0 \,.
\eeq
From \rf{man02-14112012-05} we learn that the ket-vector
$\psik$ \rf{man02-14112012-04-a} is degree-$s$ homogeneous polynomial
in the oscillators $\alpha^a$, $\zeta$. Using \rf{man02-14112012-01}, \rf{man02-14112012-02}, we introduce the following ket-vectors
\be \label{man02-17112012-01}
|\psi_\u\rangle  = \sum_{s'=0}^s\frac{\zeta^{s-s'}}{\sqrt{(s-s')!}}
|\psi_\u^{s'}\rangle \,,
\qquad
|\psi_\d\rangle  = \sum_{s'=0}^s\frac{\zeta^{s-s'}}{\sqrt{(s-s')!}}
|\psi_\d^{s'}\rangle \,.
\ee
From \rf{man02-14112012-06}, we learn that the
ket-vector $|\psi_\u\rangle$ \rf{man02-17112012-01} is degree-$k_s$
homogeneous polynomial in the oscillators $\zeta$, $\upsilon^\oplussm$,
$\upsilon^\ominussm$, while, from \rf{man02-14112012-07}, we learn that ket-vector
$|\psi_\d\rangle$ \rf{man02-17112012-01} is degree-$(k_s-1)$
homogeneous polynomial in the oscillators $\zeta$,
$\upsilon^\oplussm$, $\upsilon^\ominussm$. In terms of the
ket-vector $\psik$, the $\gamma$ triple-tracelessness constraint
\rf{gammatriple} takes the form%
\footnote{ In this paper, we adapt the formulation in terms of the $\gamma$
triple-traceless fermionic fields in Ref.\cite{Fang:1978wz}.
For the use of unconstrained gauge fields see Refs.\cite{Campoleoni:2009gs,Francia:2002aa}.}
\be \label{man02-14112012-08}
\gaalb \bar\alpha^2  \psik = 0 \,.
\ee

\subsection{ Lagrangian and gauge symmetries }\label{gaugsym}

We are now ready to discuss gauge invariant Lagrangian. Lagrangian
of conformal spin-$(s+\half)$ fermionic field we found takes the
form
\beq
\label{man02-14112012-09} && \hspace{-1cm} {\rm i} \LL =   \psibr E
\psik\,, \hspace{1cm}  E \equiv E_\smone + E_\smzero\,,
\\
\label{man02-14112012-11} && E_\smone \equiv
\parline  - \alpha\partial\gamma\bar\alpha -
\gamma\alpha\bar\alpha\partial + \gamma\alpha
\parline\gamma\bar\alpha + \frac{1}{2}\gamma\alpha\alpha\partial\bar\alpha^2
+ \frac{1}{2}\alpha^2\gamma\bar\alpha\bar\alpha\partial -
\frac{1}{4}\alpha^2\parline\bar\alpha^2\,, \ \ \ \
\\
\label{man02-14112012-12} && E_\smzero = (1 -
\gamma\alpha\gamma\bar\alpha - \frac{1}{4}\alpha^2\bar\alpha^2)
g^\Gammasm + (\gamma\alpha - \half\alpha^2\gamma\bar\alpha)\gb
+ (\gamma\bar\alpha -\half \gamma\alpha\bar\alpha^2) g\,,
\\
\label{man02-14112012-14} && g^\Gammasm = g_\zeta^\Gammasm
(\upsilon^\oplussm \sigma_+ + \bar\upsilon^\oplussm \sigma_-)\,,
\\
\label{man02-14112012-15} && g = \zeta\bar\upsilon^\oplussm g_\zeta
\,,\qquad \gb = - g_\zeta \upsilon^\oplussm \bar\zeta\,,
\\
\label{man02-14112012-16} && g_\zeta^\Gammasm  \equiv
\frac{2s+ d -2}{2s + d - 2 - 2N_\zeta}\,,  \qquad g_\zeta \equiv \Bigl(\frac{2s+ d
-3 - N_\zeta}{2s + d - 4 - 2N_\zeta}\Bigr)^{1/2} \,.
\eeq
Note that the $E_\smone$ \rf{man02-14112012-11} is the well known
first-order Fang-Fronsdal operator represented in terms of the
oscillators. The bra-vector $\psibr$ is defined according the rule
$\psibr \equiv (\psik)^\dagger \gamma^0$.

We now proceed with the discussion of gauge symmetries of Lagrangian
\rf{man02-14112012-09}. First, we introduce
gauge transformation parameters given by:
\beq
\label{man02-14112012-18} && \hspace{-0.5cm} \xi_{k'-1}^{a_1\ldots
a_{s'}} \,,\hspace{1.7cm} s'=0,1,\ldots,s-1\,, \hspace{3cm} k' \in
[k_{s'}+1]_2;
\\
\label{man02-14112012-19} && \hspace{-0.5cm} \xi_{k'-1}^{a_1\ldots
a_{s'}} \,,\hspace{1.7cm} s'=0,1,\ldots,s-1\,, \hspace{3cm} k' \in
[k_{s'}]_2;
\eeq
where the spinor indices of the fields $\xi_{k'-1}^{a_1\ldots
a_{s'}}$ are implicit. The parameters
$\xi_{k'-1}^{a_1\ldots a_{s'}}$ are non-chiral, spinor,
vector-spinor, and tensor-spinor fields of the Lorentz algebra
$so(d-1,1)$.

Gauge transformation parameters appearing in
\rf{man02-14112012-18}, \rf{man02-14112012-19} can alternatively be represented as
\beq
\label{man02-14112012-20}  &&\xi_{k'-1}^{a_1\ldots a_{s-1}}\,,
\hspace{3.2cm} k'\in [k_s]_2\,;
\\
&&\xi_{k'-1}^{a_1\ldots a_{s-1}}\,, \hspace{3.2cm} k'\in
[k_s-1]_2\,;
\\
&&\xi_{k'-1}^{a_1\ldots a_{s-2}}\,, \hspace{3.2cm} k' \in
[k_{s-1}]_2\,;
\\
&&\xi_{k'-1}^{a_1\ldots a_{s-2}}\,, \hspace{3.2cm} k' \in
[k_{s-1}-1]_2\,;
\\
&&\ldots\ldots \hspace{3.4cm} \ldots \ldots\ldots
\nonumber\\[-10pt]
&&\ldots\ldots \hspace{3.4cm} \ldots \ldots\ldots
\nonumber\\
&& \xi_{k'-1}^a\,, \hspace{3.8cm} k' \in [k_2]_2\,;
\\
&& \xi_{k'-1}^a\,, \hspace{3.8cm} k' \in [k_2-1]_2\,;
\\
\label{man02-14112012-28}  && \xi_{k'-1}\,, \hspace{3.8cm} k' \in
[k_1]_2\,;
\\
\label{man02-14112012-29}  && \xi_{k'-1}\,, \hspace{3.8cm} k' \in
[k_1-1]_2\,;
\eeq

We note that\\
\ibf) In \rf{man02-14112012-18},\rf{man02-14112012-19} the fields
$\xi_{k'-1}$ and $\xi_{k'-1}^a$ are the respective non-chiral spinor
and vector-spinor fields of the Lorentz algebra, while the fields
$\xi_{k'-1}^{a_1\ldots a_{s'}}$, $s'>1$, are rank-$s'$ totally
symmetric non-chiral tensor-spinor fields of the Lorentz algebra
$so(d-1,1)$.
\\
\iibf) The vector-spinor fields $\xi_{k'-1}^a$ and tensor-spinor
fields $\xi_{k'-1}^{a_1\ldots a_{s'}}$ with $s'\geq 1 $ satisfy the
$\gamma$-tracelessness constraint
\be \label{man02-14112012-30}
\gamma^a\xi_{k'-1}^{aa_2\ldots a_{s'}}=0\,, \qquad s'\geq 1\,.
\ee
\iiibf) The conformal dimension of the gauge transformation
parameter $\xi_{k'-1}^{a_1\ldots a_{s'}}$ is given by
\be \label{man02-15112012-01}
\Delta(\xi_{k'-1}^{a_1\ldots a_{s'}}) = \frac{d-1}{2} + k'-1\,.
\ee

Second, we collect the gauge transformation parameters into
a ket-vector $\xik$ given by
\be \label{man02-15112012-02}
\xik = \sum_{s'=0}^{s-1}\frac{\zeta^{s-1-s'}}{\sqrt{(s-1-s')!}}
|\xi^{s'}\rangle \,, \hspace{1cm}
|\xi^{s'}\rangle  = \left(
\begin{array}{l}
|\xi_\u^{s'}\rangle
\\[7pt]
|\xi_\d^{s'}\rangle \end{array} \right),
\ee

\beq
\label{man02-15112012-03u} && |\xi_\u^{s'}\rangle  \equiv \sum_{k'
\in [k_{s'}+1]_2} \frac{
(\upsilon^\ominussm)^{^{\frac{k_{s'}+1+k'}{2}}}
(\upsilon^\oplussm)^{^{\frac{k_{s'}+1-k'}{2}}} }{
s'!(\frac{k_{s'}+1+k'}{2})!}\, \alpha^{a_1} \ldots  \alpha^{a_{s'}}
\xi_{ k'-1}^{a_1\ldots a_{s'}}|0\rangle\,,
\\
\label{man02-15112012-03d}&& |\xi_\d^{s'}\rangle \equiv \sum_{k' \in
[k_{s'}]_2} \frac{ (\upsilon^\ominussm)^{^{\frac{k_{s'}+k'}{2}}}
(\upsilon^\oplussm)^{^{\frac{k_{s'}-k'}{2}}} }{
s'!(\frac{k_{s'}+k'}{2})!}\, \alpha^{a_1} \ldots  \alpha^{a_{s'}}
\xi_{k'-1}^{a_1\ldots a_{s'}}|0\rangle\,.
\eeq
We note that that ket-vector appearing in \rf{man02-15112012-02} satisfies
algebraic constraints given by
\beq
\label{man02-15112012-04} && (N_\alpha + N_\zeta - s +1) \xik = 0
\,,
\\
\label{man02-15112012-05} && (N_\zeta + N_\upsilon - k_s )\pi_+ \xik
= 0 \,,
\\
\label{man02-15112012-06} && (N_\zeta + N_\upsilon  - k_s+1)\pi_-
\xik =0 \,,
\eeq
where $k_{s'}$ is defined in \rf{man02-12112012-03}. Algebraic
constraints \rf{man02-15112012-04} imply that ket-vector $\xik$
\rf{man02-15112012-02} is degree-$(s-1)$ homogeneous polynomial in
the oscillators $\alpha^a$, $\zeta$. Introducing the notation
\be \label{man02-17112012-02}
|\xi_\u\rangle  =
\sum_{s'=0}^{s-1}\frac{\zeta^{s-1-s'}}{\sqrt{(s-1-s')!}}
|\xi_\u^{s'}\rangle \,,
\qquad |\xi_\d \rangle =
\sum_{s'=0}^{s-1}\frac{\zeta^{s-1-s'}}{\sqrt{(s-1-s')!}}
|\xi_\d^{s'}\rangle \,,
\ee
we note that algebraic constraint \rf{man02-15112012-05} implies
that ket-vector $|\xi_\u\rangle$ \rf{man02-17112012-02} is
degree-$k_s$ homogeneous polynomial in the oscillators $\zeta$,
$\upsilon^\ominussm$, $\upsilon^\oplussm$, while, from algebraic
constraint \rf{man02-15112012-06}, we learn that ket-vector
$|\xi_\d\rangle$ \rf{man02-17112012-02} is degree-$(k_s-1)$
homogeneous polynomial in the oscillators $\zeta$,
$\upsilon^\ominussm$, $\upsilon^\oplussm$. In terms of the
ket-vector $\xik$, $\gamma$-tracelessness constraint
\rf{man02-14112012-30} takes the form
\be \label{man02-15112012-07}
\gaalb \xik=0 \,.
\ee

Third, we note that gauge transformations can be presented in terms of $\psik$
and $\xik$. Namely, the gauge transformations take the form
\be
\label{man02-15112012-08}  \delta \psik = G\xik\,, \hspace{1.3cm} G \equiv \alpha\partial
-  g + \gaal \frac{1}{2N_\alpha+d-2} g^\Gammasm - \alpha^2
\frac{1}{2N_\alpha+d} \gb \,,
\ee
where operators $g^\Gammasm$, $g$, $\gb$ are defined in
\rf{man02-14112012-14}-\rf{man02-14112012-16}.

{\bf Chiral conformal fermionic fields}. In the above discussion, we
considered conformal {\it non-chiral} Dirac fermionic fields
\rf{man02-14112012-04-a}. Extension of our discussion to the case of
conformal {\it chiral} fermionic fields is straightforward. To this
end we introduce matrix $\Gamma_*$ defined as
\beq
\label{man02-17112012-03} && \Gamma_* \equiv \gamma_* \sigma_3 \,,
\qquad \Gamma_*^2 = 1\,, \qquad \Gamma_*^\dagger = \Gamma_*\,,
\\
\label{man02-17112012-04} && \gamma_* \equiv \epsilon \gamma^0
\gamma^1 \ldots \gamma^{d-1} \,, \qquad \quad
\gamma_*^2 =1\,, \qquad \gamma_*^\dagger = \gamma_*\,,\qquad \epsilon^2 = (-)^{(d-2)/2}\,, \qquad
\eeq
where $\sigma_3$ is defined in \rf{manold-03112011-01}. Now we
introduce chiral ket-vectors $|\psi_\pm\rangle$ defined as
\beq
\label{man02-17112012-05} &&  |\psi_\pm \rangle = \Pi_\pm \psik\,,
\qquad  \Pi_\pm \equiv \half (1\pm \Gamma_*)\,.
\eeq
We verify that the matrix $\Gamma_*$ anticommutes with
operator $E$ \rf{man02-14112012-09} entering Lagrangian
\rf{man02-14112012-09},
\be \label{man02-17112012-06}
\{\Gamma_*, E\} = 0 \,.
\ee
Using \rf{man02-17112012-06} and the relations
$\{\Gamma_*,\gamma^0\}= 0$, $\Pi_\pm^\dagger = \Pi_\pm$, we see that
Lagrangian \rf{man02-14112012-09} is decomposed as
\be \label{man02-17112012-07}
\LL = \LL_+ + \LL_- \qquad {\rm i} \LL_\pm = \langle \psi_\pm | E |
\psi_\pm \rangle \,.
\ee
The Lagrangian $\LL_+$ describes positive chirality conformal
fermionic field $|\psi_+\rangle$, while the Lagrangian $\LL_-$
describes negative chirality conformal fermionic field
$|\psi_-\rangle$. Note also that the projectors $\Pi_\pm$ are
commuting with operator $G$ \rf{man02-15112012-08}, $[\Pi_\pm,G] = 0$.
Taking this into account, we see that the Lagrangians $\LL_+$ and
$\LL_-$ are invariant under the respective gauge transformations
\be
\delta |\psi_+\rangle = G|\xi_+\rangle \,, \qquad \delta
|\psi_-\rangle = G|\xi_-\rangle\,,\qquad  |\xi_\pm \rangle \equiv
\Pi_\pm \xik\,.
\ee

\newsection{Reailzation of conformal symmetries
} \label{realcondsymmetr}

In order to complete our formulation of spin-$(s+\half)$
conformal fermionic field we should describe a realization of symmetries of the conformal
algebra on a space of the ket-vector $\psik$ \rf{man02-14112012-04-a}. We now present the realization we obtained.

Conformal symmetries for $R^{d-1,1}$ space are described by the $so(d,2)$ algebra. This algebra consists of translation generators $P^a$, Lorentz rotation generators $J^{ab}$ which span $so(d-1,1)$ algebra, dilatation generator $D$, and conformal
boost generators $K^a$.  We use the following commutators of the conformal algebra:%
\footnote{ Note that in our approach, only $so(d-1,1)$ symmetries
are realized manifestly. The $so(d,2)$ symmetries of conformal
fields could be realized manifestly by using ambient space approach
(see, e.g., Refs.\cite{Bekaert:2009fg}-\cite{Fotopoulos:2006ci}.)}
\beq
\label{ppkk}
&& {}[D,P^a]=-P^a\,, \hspace{2.2cm}  {}[P^a,J^{bc}]=\eta^{ab}P^c
-\eta^{ac}P^b \,,
\\
&& [D,K^a]=K^a\,, \hspace{2.4cm} [K^a,J^{bc}]=\eta^{ab}K^c -
\eta^{ac}K^b\,,
\\
\label{pkjj} &&  [P^a,K^b]=\eta^{ab}D-J^{ab}\,, \qquad [J^{ab},J^{ce}]=\eta^{bc}J^{ae}+3\hbox{ terms} \,.
\eeq

Let $\psik$ stands for fermionic field which propagate in $R^{d-1,1}$. Action for the free field $\psik$ should be invariant under the transformations
\be
\label{man-14022012-01} \delta_G \psik  = G_\diff \psik \,,
\ee
where the realization of the generators of the conformal algebra $so(d,2)$  in
terms of differential operators $ G_\diff$ acting on the ket-vector $\psik$ is given by
\beq
\label{man02-17112012-10} && P^a = \partial^a \,, \hspace{1.6cm}  J^{ab} = x^a\partial^b - x^b\partial^a
+ M^{ab}\,,
\\
\label{man02-17112012-12} && D = x\partial  + \Delta\,, \qquad K^a = K_{\Delta,M}^a + R^a\,,
\\
\label{man02-17112012-15} && \hspace{3.4cm} K_{\Delta,M}^a \equiv
-\frac{1}{2}x^2\partial^a + x^a D + M^{ab}x^b\,.
\eeq
In expressions \rf{man02-17112012-10}-\rf{man02-17112012-15}, the quantities $\Delta$  and $M^{ab}$ stand for the respective operator of conformal dimension and operator of the Lorentz algebra spin. An operator $R^a$ appearing in \rf{man02-17112012-12} depends on derivatives with respect to
space-time coordinates and does not depend on space-time coordinates
$x^a$, $[P^a,R^b]=0$.%
\footnote{For conformal currents and shadow fields
considered in Refs.\cite{Metsaev:2008fs,Metsaev:2010zu,Metsaev:2011uy},
the operator $R^a$ is independent of the derivatives.}
The operator $M^{ab}$ is known for arbitrary spin conformal fermionic field. We note that, in higher-derivative approach, the operator $R^a$ is trivial, while,
in our first-derivative formulation, the operator, in general, $R^a$ is
non-trivial. This is to say that, in our first-derivative approach, the complete formulation of conformal fields  requires finding, among other things, the operator $R^a$.

Thus, all that remains, is to find the operators $M^{ab}$, $\Delta$, and $R^a$.
For the case of totally symmetric arbitrary spin-$(s+\half)$ conformal fermionic field, the
spin operator $M^{ab}$ and the conformal dimension operator $\Delta$ are
given by
\beq
\label{man02-15112012-10} && M^{ab} = \alpha^a \bar\alpha^b -
\alpha^b \bar\alpha^a  + \half \gamma^{ab}\,,
\\
\label{man02-15112012-11} && \Delta  =  \frac{d-1}{2}+\Delta'\,,
\qquad \Delta' \equiv N_{\upsilon^\ominussm} - N_{\upsilon^\oplussm}
\,.
\eeq
Obviously, expression for the conformal dimension operator $\Delta$ given in
\rf{man02-15112012-11} can be read from the relations given in \rf{condimarbspi01}.
Realization of the operator $R^a$ on fermionic ket-vector $\psik$ we found takes the form
\beq
\label{man02-16112012-01} && R^a  = r_\smzero^a + r_\smone^a +
R_\smG^a + R_\smE^a\,,
\\
\label{man02-16112012-02} && r_\smzero^a  = r_{0,1}^\Gammasm \GG^a +
r_{0,1}\bar\alpha^a + \rb_{0,1}\AA^a\,,
\qquad
r_\smone^a =  r_{1,1} \partial^a\,,
\\
\label{man02-16112012-03} && R_\smG^a  =   G  r_\smG^a\,,
\\
\label{man02-16112012-04} && R_\smE^a  =    r_\smE^a E \,,
\\
\label{man02-16112012-05}  && \hspace{1cm} r_{0,1}^\Gammasm =
g_\zeta^\Gammasm ( \bar\upsilon^\ominussm \sigma_- - \upsilon^\ominussm \sigma_+)\,,
\\
\label{man02-16112012-06}  && \hspace{1cm} r_{0,1} = 2 \zeta g_\zeta
\bar\upsilon^\ominussm\,,
\qquad
\rb_{0,1} = - 2 \upsilon^\ominussm g_\zeta \bar\zeta\,,
\qquad
r_{1,1} = -2\upsilon^\ominussm \bar\upsilon^\ominussm \,,
\eeq
\beq
\label{man02-16112012-07} r_\Gsm^a & = & r_{\Gsm,1} \GG_1^a
\Pi^\smponethree + r_{\Gsm ,2}\GG_1^a \Pi^\smponethree \gaalb  +
r_{\Gsm ,3} \GG_1^a \bar\alpha^2
\nonumber\\
&  + & r_{\Gsm,4} \Pi^\smponethree \bar\alpha^a + r_{\Gsm ,5}
\Pi^\smponethree \gaalb \bar\alpha^a + r_{\Gsm ,6}
\bar\alpha^a\bar\alpha^2
\nonumber\\
& + & r_{\Gsm,7} \AA_1^a \Pi^\smponethree  + r_{\Gsm,8} \AA_1^a
\Pi^\smponethree \gaalb  + r_{\Gsm,9} \AA_1^a \bar\alpha^2\,,
\\
\label{man02-16112012-08} r_\Esm^a & = &  r_{\Esm,1}\gamma^a
\Pi^\smponetwo + r_{\Esm,2} \gaal \gamma^a \gaalb + r_{\Esm,3}
\alpha^2 (\GG_1^a \Pi^\smponethree + \Pi^\smponethree
\bar\GG_1^a)\bar\alpha^2
\nonumber\\
& + & \Bigl( r_{\Esm,4} \gaal \GG_1^a \Pi^\smponethree + r_{\Esm,5}
\alpha^2 \GG_1^a \Pi^\smponethree \gaalb + r_{\Esm,6} \alpha^2
\GG_1^a \Pi^\smponethree
\nonumber\\
& +& r_{\Esm,7} \AA_1^a \Pi^\smponethree \gaalb + r_{\Esm,8} \gaal
\AA_1^a\Pi^\smponethree \bar\alpha^2 + r_{\Esm,9} \AA^a
\nonumber\\
& +& r_{\Esm,10} \gaal \AA_1^a\Pi^\smponethree\gaalb +
r_{\Esm,11}\alpha^2 \Pi^\smponethree \bar\AA_1^a + r_{\Esm,12}
\alpha^2 \AA_1^a\Pi^\smponethree \bar\alpha^2
\nonumber\\
&  + & r_{\Esm,13} \gaal \AA_1^a \Pi^\smponethree + r_{\Esm,14}
\alpha^2 \AA_1^a\Pi^\smponethree \gaalb + r_{\Esm,15}\alpha^2
\AA_1^a\Pi^\smponethree - \widehat{h.c.}\Bigr)\,,
\eeq
\beq
\label{man02-19112012-01} && r_{\Gsm,n} =  \upsilon^\ominussm
\rwt_{\Gsm,n+} \bar\upsilon^\ominussm \pi_+ + \upsilon^\ominussm
\rwt_{\Gsm,n-}\bar\upsilon^\ominussm \pi_-   \,,\hspace{2.5cm}
n=2,4,9\,;
\\
&& r_{\Gsm,n} =  (\upsilon^\ominussm \upsilon^\ominussm \rwt_{\Gsm,n+}
 \sigma_+ + \upsilon^\ominussm
\rwt_{\Gsm,n-}\bar\upsilon^\ominussm \sigma_-)\bar\zeta
\,,\hspace{2cm} n=1,8\,;
\\
&& r_{\Gsm,n} =  \zeta (\upsilon^\ominussm \rwt_{\Gsm,n+}
\bar\upsilon^\ominussm \sigma_+ +
\rwt_{\Gsm,n-}\bar\upsilon^\ominussm\bar\upsilon^\ominussm \sigma_-)
\,,\hspace{2cm} n=3,5\,;
\\
&& r_{\Gsm,n} =  (\upsilon^\ominussm \upsilon^\ominussm \rwt_{\Gsm,n+}
\pi_+ + \upsilon^\ominussm \upsilon^\ominussm \rwt_{\Gsm,n-}
\pi_-)\bar\zeta^2 \,,\hspace{2cm} n=7\,;
\\
&& r_{\Gsm,n} =  \zeta^2 (\rwt_{\Gsm,n+} \bar\upsilon^\ominussm
\bar\upsilon^\ominussm \pi_+ +
\rwt_{\Gsm,n-}\bar\upsilon^\ominussm\bar\upsilon^\ominussm \pi_-)
\,,\hspace{2cm} n=6\,;
\\
&& r_{\Esm,n} =   \upsilon^\ominussm \rwt_{\Esm,n+}
\bar\upsilon^\ominussm \pi_+ + \upsilon^\ominussm
\rwt_{\Esm,n-}\bar\upsilon^\ominussm \pi_-  \,,\qquad \qquad \qquad
n=1,2,3,7,8\,;
\\
&& r_{\Esm,n} = (\upsilon^\ominussm \upsilon^\ominussm \rwt_{\Esm,n+}
\sigma_+  + \upsilon^\ominussm \rwt_{\Esm,n-} \bar\upsilon^\ominussm
\sigma_-)\bar\zeta \,,\hspace{2cm} n=4,5,9,10,11,12\,;
\\
&& r_{\Esm,n} = (\upsilon^\ominussm \upsilon^\ominussm \rwt_{\Esm,n+}
\pi_+  + \upsilon^\ominussm \upsilon^\ominussm
\rwt_{\Esm,n-}\pi_-)\bar\zeta^2 \,, \hspace{2cm} n=6,13,14\,;
\\
&& r_{\Esm,n} = (\upsilon^\ominussm
\upsilon^\ominussm\upsilon^\ominussm\upsilon^\oplussm
\rwt_{\Esm,n+}\sigma_+  +  \upsilon^\ominussm \upsilon^\ominussm
\rwt_{\Esm,n-}\sigma_-)\bar\zeta^3  \,, \hspace{1.1cm} n=15\,;
\\
\label{man02-19112012-15} && r_{\Esm,n}^\dagger = r_{\Esm,n},\,,
\qquad \hbox{ for } \ n=1,2,3\,,
\eeq
where the quantities $G$, $E$ \rf{man02-16112012-03},\rf{man02-16112012-04} are defined in \rf{man02-15112012-08},\rf{man02-14112012-09}, while the quantities $g_\zeta^\Gammasm$, $g_\zeta$  \rf{man02-16112012-05}, \rf{man02-16112012-06} are defined in \rf{man02-14112012-16}. In \rf{man02-19112012-01}-\rf{man02-19112012-15}, the quantities
$\rwt_{\smG,n\pm}$ and $\rwt_{\smE,n\pm}$ remain to be arbitrary functions of
the operators $N_\zeta$, $\Delta'$.

The following remarks are in order.

\noindent {\bf i}) The operators $r_\smzero^a$ and $r_\smone^a$
\rf{man02-16112012-02} entering the operator $R^a$ are fixed unambiguously, while the operators $R_\smG^a$, $R_\smE^a$ remain to be arbitrary because of the
$\rwt_{\smG,n\pm}$, $\rwt_{\smE,n\pm}$ are arbitrary functions of
the $N_\zeta$, $\Delta'$.
The reason for arbitrariness governed by the operator $R_\smG^a$ is clear: the $so(d,2)$ algebra transformations of all gauge fields are defined by module of
gauge transformations. As the operator $R_\smG^a$
\rf{man02-16112012-03} is proportional to the operator of gauge
transformation $G$, the action of the operator $R_\smG^a$ on ket-vector $\psik$
takes the form of gauge transformation with a gauge
transformation parameter equal to $r_\smG^a\psik$. The reason for arbitrariness governed by the operator $R_\smE^a$ is also clear: the $so(d,2)$ algebra transformations of fermionic fields are defined by module of $\tau E\psik$, where $\tau$ is an arbitrary operator that satisfies the condition $\tau^\dagger =
\gamma^0 \tau\gamma^0$. It is easy to see that operator $r_\Esm^a$
given in \rf{man02-16112012-08} satisfies this condition.

\noindent {\bf ii}) Considering $R_\smG^a=0$, $R_\smE^a=0$, we verify the commutator $[K^a,K^b]=0$.

\noindent {\bf iii}) Considering $R_\smG^a \ne 0 $, $R_\smE^a \ne 0 $,  we find the following expression for commutator $[K^a,K^b]$:
\beq
\label{man02-16112012-09} && [K^a,K^b] =  W^{ab}\,,
\\
\label{man02-16112012-10} && \hspace{2cm} W^{ab} \equiv G
r_\smG^{ab} + r_\smE^{ab} E + G r_\smGE^{ab} E\,,
\\
&& \hspace{2cm} r_\smG^{ab} \equiv r^a r_\smG^b + r_\smG^a r^b +
r_\smG^a G r_\smG^b - (a \leftrightarrow b)\,,
\\
\label{man02-16112012-10a} && \hspace{2cm} r_\smE^{ab} \equiv r^a
r_\smE^b + r_\smE^b r^{a\hat{\dagger}} + r_\smE^a E r_\smE^b - (a
\leftrightarrow b)\,, \qquad r^{a\hat{\dagger}} \equiv -\gamma^0
r^{a\dagger}\gamma^0\,, \qquad
\\
&& \hspace{2cm} r_\smGE^{ab} \equiv r_\smG^a r_\smE^b - (a
\leftrightarrow b)\,,
\qquad  r^a \equiv r_\smzero^a + r_\smone^a\,.
\eeq
From \rf{man02-16112012-09}, we learn that the commutator $[K^a,K^b]$
is proportional to the operator of gauge transformations $G$ and to
operator $E$ as it should be for fermionic gauge fields.

\noindent {\bf iv}) It is easy to see that the operator $R^a$
\rf{man02-16112012-01} is commuting with the projectors $\Pi_\pm$
\rf{man02-17112012-05} which enter positive and negative chirality
conformal fields $|\psi_\pm\rangle$,
\be \label{man02-17112012-09}
[\Pi_\pm, R^a ] = 0 \,.
\ee
Using \rf{man02-17112012-09}, we make sure that the projectors
$\Pi_\pm$ are commuting with all generators of conformal algebra
\rf{man02-17112012-10},\rf{man02-17112012-12}. This implies that
Lagrangians $\LL_+$ and $\LL_-$ \rf{man02-17112012-07} for the
respective positive and negative chirality conformal fields are
invariant under conformal algebra transformations.

To summarize we note that the Lagrangian, gauge transformations,
and the operator $R^a$ of the conformal fermionic fields are fixed unambiguously by the following three requirements.%
\footnote{ Discussion of uniqueness of the higher-derivative formulation for interacting spin-2 conformal field theory may be found in Ref.\cite{Boulanger:2001he}.}
\\
\ibf) Lagrangian and gauge transformations of fermionic conformal fields should not consist higher than first order terms in derivatives.
\\
\iibf) the operator $R^a$ entering conformal boost generators should not consist higher than first order terms in derivatives;\\
\iiibf) Action of the fermionic gauge field should be invariant under conformal algebra transformations and gauge transformations.

These three requirements allow us to fix the Lagrangian and gauge
transformations unambiguously. The operator $R^a$ is also fixed unambiguously
by module of the gauge transformation operator $G$ and the operator $E$ (as
it should be for fermionic gauge fields).

\newsection{ \large On-shell degrees of freedom of conformal
field }\label{dof}

For $d=4$ and $s\geq 1$, on-shell D.o.F of the
spin-$(s+\half)$ conformal fermionic field were found in
Ref.\cite{Fradkin:1985am}. Decomposition of on-shell D.o.F into
irreps of the $so(2)$ algebra was considered only for the case of spin-$\frac{3}{2}$ conformal field in Ref.\cite{Lee:1982cp}. For arbitrary values
$s$, $s\geq 1$, and arbitrary dimension of space, $d > 4$, on-shell
D.o.F of the spin-$(s+\half)$ conformal fermionic field have not
been discussed so far in the literature. In this section, first,
we present our result for on-shell D.o.F of the totally symmetric
arbitrary spin-$(s+\half)$ conformal fermionic field in
$R^{d-1,1}$ space. Second, we present our results for
the decomposition of the on-shell D.o.F into irreps of the $so(d-2)$
algebra for the case of arbitrary values of $s$ and $d$.

In order to find on-shell D.o.F of the conformal fermionic field we
use the light-cone gauge. This is to say that we use fields that transform as non-chiral representations
of the $so(d-2)$ algebra
and decompose the on-shell D.o.F of conformal fermionic field into such non-chiral representations.%
\footnote{Alternative methods for counting on-shell
D.o.F were discussed in Refs.\cite{Lee:1982cp,Buchbinder:1987vp}.}
We find that complex-valued on-shell D.o.F of the
totally symmetric spin-$(s+\half)$ conformal non-chiral fermionic
field in $d$-dimensional space, $d\geq 4$, are described by the
following set of non-chiral half-integer fields of the $so(d-2)$
algebra:
\beq
&& \hspace{-0.5cm} \psi_{k'}^{i_1\ldots i_{s'}} \,,\hspace{1.7cm}
s'=0,1,\ldots,s\,, \hspace{5.3cm} k' \in [k_{s'}]_2;
\nonumber\\[-10pt]
&& \label{man02-16112012-18}
\\[-10pt]
&& \hspace{-0.5cm} \psi_{k'}^{i_1\ldots i_{s'}}\,, \hspace{1.7cm}
s'= \left\{\begin{array}{l}
0,1,\ldots,s;\qquad  \hbox{for }\  d \geq 6;
\\[3pt]
1,2,\ldots,s;\qquad  \hbox{for }d = 4;
\end{array}\right.
\hspace{2cm} k' \in  [k_{s'}-1]_2\,;\qquad
\nonumber
\eeq
where $k_{s'}$ is defined in \rf{man02-12112012-03} and vector
indices of the $so(d-2)$ algebra take the following values $i=1,2,\ldots,d-2$. In
\rf{man02-16112012-18}, the fields $\psi_{k'}$ and $\psi_{k'}^i$ are
the respective non-chiral spinor and vector-spinor fields of the
$so(d-2)$ algebra, while the fields $\psi_{k'}^{i_1\ldots i_{s'}}$,
$s'\geq 2$, are rank-$s'$ totally symmetric tensor-spinor fields of
the $so(d-2)$ algebra. Fields $\psi_{k'}^{i_1\ldots i_{s'}}$ with
$s'\geq 1$ are $\gamma$-traceless,
\be \label{man02-16112012-19}
\gamma^i\psi_{k'}^{ii_2\ldots i_{s'}}=0\,, \qquad s'\geq 1\,.
\ee
In view of \rf{man02-16112012-19}, the vector-spinor and tensor-spinor fields
$\psi_{k'}^{i_1\ldots i_{s'}}$ transform as non-chiral irreps of the
$so(d-2)$ algebra. Obviously, our conformal fermionic field
is related to non-unitary
representation of the conformal algebra $so(d,2)$.%
\footnote{ Detailed study of representations of (super)conformal
algebras may be found, e.g., in Refs.\cite{Evans}-\cite{Dobrev:1985qv}.}

For $d\geq 6$, field content \rf{man02-16112012-18}
can alternatively be represented as
\beq
\label{man02-16112012-20} &&\psi_{k'}^{i_1\ldots i_s}\,,
\hspace{3cm} k'\in [k_s]_2\,;
\\
&&\psi_{k'}^{i_1\ldots i_s}\,, \hspace{3cm} k'\in [k_s-1]_2\,;
\\
&&\psi_{k'}^{i_1\ldots i_{s-1}}\,, \hspace{2.7cm} k' \in
[k_{s-1}]_2\,;
\\
&&\psi_{k'}^{i_1\ldots i_{s-1}}\,, \hspace{2.7cm} k' \in
[k_{s-1}-1]_2\,;
\\[-5pt]
&&\ldots\ldots \hspace{3.4cm} \ldots \ldots\ldots
\nonumber\\[-10pt]
&&\ldots\ldots \hspace{3.4cm} \ldots \ldots\ldots
\nonumber\\[-5pt]
&& \psi_{k'}^i\,, \hspace{3.8cm} k' \in [k_1]_2\,;
\\
&& \psi_{k'}^i\,, \hspace{3.8cm} k' \in [k_1-1]_2\,;
\\
\label{man02-16112012-29} && \psi_{k'}\,, \hspace{3.8cm} k' \in
[k_0]_2\,;
\\
\label{man02-16112012-30} && \psi_{k'}\,, \hspace{3.8cm} k' \in
[k_0-1]_2\,.
\eeq
For $d=4$, field content \rf{man02-16112012-18} is given in
\rf{man02-16112012-20}-\rf{man02-16112012-29}. This is to say that
spinor fields in \rf{man02-16112012-30} enter field content only for
$d\geq 6$.

Number of the complex-valued on-shell D.o.F of non-chiral
fermionic fields shown in \rf{man02-16112012-18} is given by

\be \label{man02-16112012-11}
\nbf = 2^{\frac{d-2}{2}}(d-3) (2s+d-2) \frac{(s+d-3)!}{s!(d-2)!}\,.
\ee
For $d=4$, relation \rf{man02-16112012-11}
gives the following expressions for $\nbf$:

\be
\label{man02-16112012-12}  \nbf\big|_{s- {\rm arbitrary};\,\,
d=4}\, = \, 2(s+1)^2 \,.
\ee

Result for $\nbf$ in \rf{man02-16112012-12} was obtained in
Ref.\cite{Fradkin:1985am}. Thus, for $d=4$ and arbitrary $s$, our result
\rf{man02-16112012-11} agrees with result in Ref.\cite{Fradkin:1985am}
and provides $\nbf$ \rf{man02-16112012-11}
for the case of arbitrary $s$ and $d$.

To summarize, our result presented in relation \rf{man02-16112012-18} provides the
decomposition of the complex-valued on-shell D.o.F into non-chiral half-integer
representations of $so(d-2)$ algebra, while our expression for $\nbf$
\rf{man02-16112012-11} gives the complex-valued number of on-shell D.o.F of non-chiral fermionic conformal fields appearing in \rf{man02-16112012-18}.%

Expression for $\nbf$ in \rf{man02-16112012-11} describes number of
the complex-valued on-shell D.o.F of non-chiral conformal field. Numbers of
complex-valued on-shell D.o.F of chiral conformal fermionic fields
$|\psi_\pm\rangle$ \rf{man02-17112012-05}, which we denote by
$\nbf_\pm$, are obtained in a obvious way

\be
\nbf_\pm = \half \nbf\,.
\ee

Finally, let us explain how $\nbf$ \rf{man02-16112012-11} is obtained from decomposition in \rf{man02-16112012-18}. By definition, $\nbf$
\rf{man02-16112012-11} is a sum of tensor-spinorial components of fermionic fields
\rf{man02-16112012-18} that subject to algebraic $\gamma$-tracelessness constraint
\rf{man02-16112012-19}. Namely, the $\nbf$ can be
represented as

\beq
\label{man02-16112012-14} && \nbf = \sum_{s'=0}^s \nbf^{s'} \,,\qquad
\nbf^{s'} = \sum_{k'\in [k_s']_2} \nbf(\psi_{k'}^{s'}) + \sum_{k'\in
[k_s'-1]_2} \nbf(\psi_{k'}^{s'})\,,
\\
\label{man02-16112012-16} && \nbf(\psi_{k'}^{s'}) = 2^{\frac{d-2}{2}}
\frac{(s'+d-4)!}{s'!(d-4)!}\,,
\eeq
where $\nbf(\psi_{k'}^{s'})$ is a number of D.o.F of rank-$s'$
$\gamma$-traceless non-chiral tensor-spinor field $\psi_{k'}^{i_1\ldots
i_{s'}}$. In other words, $\nbf(\psi_{k'}^{s'})$ is a dimension of
the rank-$s'$ $\gamma$-traceless non-chiral tensor-spinor field of
the $so(d-2)$ algebra. Using expression given in \rf{man02-16112012-16}, we find that relation
for $\nbf^{s'}$ \rf{man02-16112012-14} leads to following expression:

\beq \label{man02-16112012-17}
\nbf^{s'}  & = &   2^{\frac{d-2}{2}}
\frac{(s'+d-4)!}{s'!(d-4)!} (2k_{s'}+1)
\nonumber\\
& = & 2^{\frac{d-2}{2}} \frac{(s'+d-4)!}{s'!(d-4)!}(2s'+d-3)\,.
\eeq
We now note that expression for $\nbf$  given in
\rf{man02-16112012-11} is obtained by substitution of $\nbf^{s'}$ \rf{man02-16112012-17} into
$\nbf$ in \rf{man02-16112012-14} and using the following relation,

\be \sum_{s'=0}^s \frac{(s' + t )!}{s'!} = \frac{(s+ t
+1)!}{(t+1)s!}\,.
\ee
%

\newsection{Conclusions}\label{conlcus}

In this paper, we generalized results in Refs.\cite{Metsaev:2007fq,Metsaev:2007rw} to the case of arbitrary spin conformal fermionic fields. The results
presented in this paper might have the following further interesting applications and
generalizations.

\noindent \ibf) BRST formulation of conformal fermionic fields.
BRST approach have fruitfully been used for
the investigation of gauge invariant formulation of massive fields (see,
e.g., Refs.\cite{Buchbinder:2005ua}). We noted in
Ref.\cite{Metsaev:2007rw}, that gauge invariant formulation of massive fields and  our formulation of conformal fields have
many common features. Recent discussion of BRST method for higher-spin massive fermionic fields may be found in Ref.\cite{Reshetnyak:2018fvd}. Therefore we think that first-derivative formulation of conformal fermionic fields in the framework of BRST method can straightforwardly be reached.  The second-derivative BRST formulation of bosonic conformal fields was obtained in Ref.\cite{Metsaev:2015yyv}.

\noindent \iibf) Interacting conformal fields theories.
We note that most of approaches to theories of interacting higher-spin fields in Refs.\cite{Fradkin:1987ks}-\cite{Buchbinder:2012iz} can straightforwardly be adapted for the case of conformal fields. In our approach to conformal fields, use of Stueckelberg fields is similar to the use of Stueckelberg fields in gauge invariant approach to massive fields. As is well known, the Stueckelberg fields provide systematical setup for the investigation of interacting massive gauge fermionic fields (see, e.g., Ref.\cite{Metsaev:2006ui}). We think therefore that application of our approach to theory of interacting conformal fermionic fields
might lead to new interesting development. It is worthwhile to mention the BRST approach also involves Stueckelberg fields and this approach turns out to be fruitful for the investigation of interacting higher-spin field theories (see, e.g.,  Refs.\cite{Bekaert:2010hp}-\cite{Polyakov:2010qs}).

\noindent \iiibf) Fermionic unconstrained conformal gauge fields. Formulations of various higher-spin field theories  in terms of unconstrained gauge fields were studied in Refs.\cite{Campoleoni:2009gs,Francia:2002aa}. We think that adaptation of those formulations for conformal fields might be useful for better understanding the first-derivative conformal fermionic fields.

\noindent \ivbf) Mixed-symmetry conformal fermionic fields. In the recent time, mixed-symmetry fields have extensively been studied in the literature (see, e.g., Ref.\cite{Bekaert:2002dt}). We think that, in view of potentially interesting application to string theory, study of the first-derivative formulation of mixed-symmetry conformal fermionic fields is important. Discussion of
higher-derivative mixed-symmetry bosonic conformal fields may be found in Ref.\cite{Vasiliev:2009ck}, while the discussion of self-dual conformal fields in the framework of the second-derivative approach may be found in Ref.\cite{Metsaev:2008ba}. Needless to say that the study of conformal fermionic fields along the lines in
Ref.\cite{RodGover:2012ib} could be also of some interest.

\bigskip
{\bf Acknowledgments}. This work was supported by RFBR Grant No.11-02-00685.

\end{document}